\documentclass[prb,twocolumn,showpacs]{revtex4}
\usepackage{amsfonts,amsmath,mathrsfs,epsfig,amsbsy,bm,verbatim}

\begin{document}
\title{Conductance beyond the Landauer limit and charge pumping in quantum wires}
\author{Jay D. Sau$^1$}
\author{Takuya Kitagawa$^{1}$}
\author{Bertrand I. Halperin$^{1}$}
\affiliation{$^1$Department of Physics, Harvard University, Cambridge, MA 02138, USA}

\date{\today}

\begin{abstract}
Periodically driven systems, which can be described by Floquet theory, have been proposed to show
 characteristic behavior that is distinct from static Hamiltonians. Floquet theory proposes to
 describe such periodically driven systems in terms of states that are indexed by a photon number in addition to the usual Hilbert space 
of the system. We propose a way to  measure directly this additional Floquet degree of freedom by the measurement
 of the DC conductance of a single channel quantum point contact. Specifically, we show that a single channel wire augmented with 
a grating structure when irradiated with microwave radiation can show a DC conductance above the limit of one conductance
 quantum set by the Landauer formula.
 Another interesting feature of the proposed system is that being non-adiabatic 
in character, it can be used to pump a strong gate-voltage dependent photo-current even with linearly polarized radiation.
\end{abstract}

\maketitle

\section{Introduction}
Harmonically time-varying Hamiltonians offer an interesting range of  phenomena such as 
the coherent destruction of tunneling,\cite{tunneling} photon-assisted tunneling,\cite{patunneling} adiabatic 
pumping \cite{adpumping,torres,brouwer} and non-adiabatic pumping.\cite{nadpumping} Recently it has been realized 
that such periodically time-dependent Hamiltonians can possibly be used to realize topological phases in 
materials, which were originally non-topological.\cite{lindner, takuya,takuya1}
Systems with periodically time-varying Hamiltonians (i.e. $H(t+\mathcal{T})=H(t)$)
 are described in terms of a set of quasi-stationary quasi-energy eigenstates $e^{i E t}\psi(t)$,
which unlike stationary states of static Hamiltonians vary periodically in time $\psi(t+\mathcal{T})=\psi(t)$
 and have energy eigenvalues $E$ that are only defined modulo $\Omega=2\pi/\mathcal{T}$, where $\mathcal{T}$
 is the period of the perturbation.\cite{sambe}
These quasi-energy eigenstates, which we will refer to as Floquet states, can be determined from a 
time-dependent Schrodinger equation
\begin{equation}
(H(t)-i\partial_t)\psi(t)=E\psi(t),\label{tdse}
\end{equation}
where the quasi-energy eigenvalue $E$ satisfies the condition $E\in[-\frac{\Omega}{2},\frac{\Omega}{2}]$.

As a special case we may consider a time-independent Hamiltonian $H(t)=H(t=0)$.
In this case, the frequency $\Omega$ is arbitrary, and the Floquet states are just 
eigenstates of the Hamiltonian. If $\varepsilon_s$ is the corresponding eigenvalue of $H$,
then the quasi-energy $E_s$ is given by  $E_s=\varepsilon_s-n_s\Omega$, where $n_s$ is the unique Floquet-index integer  
needed to ensure that $E_s\in[-\frac{\Omega}{2},\frac{\Omega}{2}]$. 
Note that we have considered a time-independent Hamiltonian for illustrative purposes only.
This case is slightly pathological since the frequency $\Omega$ is not uniquely defined for a
 time-independent Hamiltonian, and in principle one can 
define the quasi-energy $E_s$ relative to any frequency $\Omega$ one chooses. 
On the other hand, the Floquet index $n_s$, defined in terms of the energy eigenvalues $\epsilon_s$
 does not apply to general periodically time-dependent Hamiltonians 
where the energy $\epsilon_s$ is not well-defined.
However,  the effectively stationary states, which are classified by an energy
 $\varepsilon$ in the time-independent case, must be classified 
by quasi-energy in a limited range $E\in[-\frac{\Omega}{2},\frac{\Omega}{2}]$ in the  
general periodically time-dependent case. Therefore in general, one expects a much larger density 
of Floquet-eigenstates, which can participate in processes associated with a specific quasi-energy eigenvalue $E$.
 The increased density of states at a fixed quasi-energy $E$ can be thought of as resulting from the fact that 
a system in a putative state with energy $\varepsilon$ can be driven into a state with energy $\varepsilon+n\Omega$ and 
back by a strong time-dependent Hamiltonian.
The expanded state space accessible to states with a fixed quasi-energy $E$, which results from the 
possibility of coherent absorption and emissions of quanta with energy $\Omega$, 
 is what is responsible for the rich phenomenology of periodically driven systems.

States characterized by fixed energy or quasi-energy play a crucial role in scattering and transport processes.
Since the density of states at a fixed quasi-energy $E$ in a periodically driven system is dramatically increased 
as compared to static systems, one expects the transport property of driven systems to be significantly different 
from static systems. Such dramatic differences have already been seen in micro-wave resistance oscillation experiments 
in two-dimensional electron gases in magnetic fields.\cite{zudov,mirlin} More specifically, if one considers the 
dispersion of a single-channel quantum wire in the Floquet representation, which is written as $E_{n,k}=\varepsilon_k-n\Omega$,
one immediately observes that the number of Floquet channels of the wire is significantly larger in the 'quasi-energy'
 Floquet representation as compared to the static representation with fixed energy $\varepsilon_k$.

 A natural question to ask is whether the increased  effective number of Floquet channels in a periodically driven one-dimensional 
quantum system modifies the transport in a direct way.
 In fact, the Landauer formula for coherent electronic transport through a static quantum system,
 \cite{landauer1,landauer2,landauer3} relates the conductivity to  the number of channels in a
 transport system.
 In particular, the conductance of an $N$ channel quantum wire of non-interacting electrons is bounded above by $N G_0$,
 where $G_0=2 e^2/h$ is the quantum of conductance. The 
bound on the conductance is reached for quantum wires which are non-interacting and are connected to reservoirs by 
perfectly transmitting contacts. The increased number of Floquet channels for driven systems
 suggests that the Floquet channels could directly manifest themselves by contributing 
to the conductance leading to a conductance excess of $G_0$ in driven one-dimensional systems, at least in principle.
 In this paper, we show that this is indeed the case, 
and a single channel periodically driven quantum wire with appropriately designed contacts can have a DC conductance 
that is in excess of the Landauer bound $G_0$.
 Moreover, for geometries which are not symmetric between the left 
and right leads, the application of a time-periodic drive will be found to result in a  DC pumping current even at zero
 voltage bias, similar to previously studied systems.~\cite{adpumping,nadpumping}
 Since our system will need to 
be driven at  a finite frequency, we will find that even a quantum wire driven by linearly polarized microwave radiation 
displays a gate-voltage dependent pumping current. This is in contrast to the adiabatic pumping 
in the limit of small driving frequency where it has been shown that a two-parameter drive such as one resulting  
 from circularly polarized radiation is necessary to drive a significant pumping current. ~\cite{brouwer}      
A similar enhancement of the conductance in strongly-interacting driven quantum wires \cite{feldman} has been 
predicted using the Bosonization approach. Moreover, it has been suggested that such an enhancement might apply to the weakly 
interaction case.\cite{feldman} Also, single-parameter non-adiabatic pumping has also been demonstrated in experiments, although 
it is attributed to a mechanism different from the one discussed in the present paper.
 \cite{kaestner} This paper provides a unified picture that is based on 
the Floquet-state interpretation for enhanced conductance for weakly interacting systems and correspondingly strongly 
gate voltage dependent pumping.

\subsubsection*{Outline}
We start by reviewing in Sec. II the scattering matrix formalism in Floquet space. Following this in Sec. III, we discuss the 
analog of the Landauer formula and the resulting conductances for a two-terminal geometry in a driven system.\cite{torres,floquet_transport}
 In this section, we 
also review how unitarity of the scattering matrices leads to the bound on the conductance obtained by Landauer and 
show how in principle this bound may be violated in systems with  time-dependent Hamiltonians.
In Secs. IV and V we discuss the details of a specific wire geometry that is calculated to support, in principle, a conductance 
in excess of the Landauer limit. Furthermore, the same structure is found to carry a pumping current.
We estimate parameters for a realization of this structure in a GaAs two-dimensional electron gas (2DEG) 
and show that the excess conductance and pumping current can be measured for experimentally reasonable parameters.
 In Sec. VI, we compare the 
pumping current obtained in our structure to the adiabatic pumping proposals using circularly, or elliptically  polarized radiation. 

\section{Floquet scattering theory}
\begin{figure}
\centering
\includegraphics[scale=0.4,angle=0]{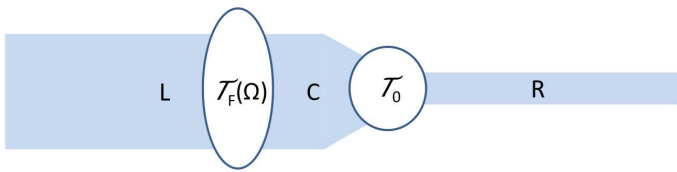}
\caption{
Schematic of conductance set-up from a 2 channel left lead $L$ to a single-channel right lead $R$ through a 2 channel 
intermediate region $C$. The transmission matrix $\mathcal{T}_0$ is static such that one of the 2 channels 
in $C$ is perfectly transmitted into $R$ while the other channel in $C$ is reflected back into $C$.
The floquet matrix $\mathcal{T}_F(\Omega)$ contains a harmonic frequency but is reflectionless and can only 
transfer electrons between one channel and the other.
}\label{fig1}
\end{figure}

Let us now discuss the problem of scattering of electrons by a harmonically time-varying potential $V$, with frequency $\Omega$, 
that is localized in a region in space $C$ (shown in Fig.~\ref{fig1}) separating a left-lead $L$ and a right lead $R$.
 Each of the leads are assumed to support multiple propagating 
channels indexed by $p$.
 The periodically time-dependent Hamiltonian for the multi-channel system 
is written as 
\begin{equation}
H_{p',p}(t)=(-\frac{\partial_x^2}{2 m}+\Delta_{p'}-\epsilon_F)\delta_{p',p}+[V_{p',p}(x)e^{i\Omega t}+h.c],\label{hpq}
\end{equation}
where $x$ is the coordinate along the wire, 
$p',p$ are channel indices, $m$ is the effective mass of the electrons, $\Delta_p$ is the confinement energy 
induced mini-gap between the various channels, $\epsilon_F$ is the Fermi-energy of the electrons in the wire
 and $V_{p',p}(x)$ are the position-dependent matrix elements 
of the potential applied in the central region. 

The solutions of the corresponding Schrodinger equation, Eq.~\ref{tdse}, 
can be written in terms of the time-independent wave-functions $\phi_{n,p}(x)$
\begin{equation}
\psi_p(x,t)=\sum_n \phi_{n,p}(x) e^{i\Omega n t},\label{floquetstates}
\end{equation}   
where the index $n$ is referred to as the Floquet index. The wave-functions   $\phi_{n,p}(x)$,
are are solutions of 
the time-independent Floquet equation,\cite{sambe} which is written as 
\begin{equation}
\sum_{n,p'} (H^{(m-n)}_{p,p'}-n\Omega)\phi_{n,p'}=E \phi_{m,p},\label{eq:floquet}
\end{equation}
where $H(t)=\sum_n H^{(n)} e^{i n\Omega t}$.
For the specific case of the quasi one-dimensional wire described by Eq.~\ref{hpq}, the components $H^{(n)}$ are written as  
\begin{align}
&H^{(0)}=\left(-\frac{\partial_x^2}{2 m}-\Delta_p\right)\delta_{p',p}\nonumber\\
&H^{(1)}=H^{(-1)\dagger}=V_{p',p}(x).\label{fullH}
\end{align}
 
 Since the time-dependent potential $V$ vanishes outside the central region $C$, the solutions in the leads  to Eq.~\ref{eq:floquet}
  can be expanded in terms of plane-wave energy eigenstates 
\begin{equation}
\phi_{n,p}(x)=e^{i k_{n,p} x}\phi_{n,p}(0)
\end{equation}
at a pre-determined quasi-energy $E$, where 
the wave-vectors $k_{n,p}$ are determined by the requirement that the Floquet energy $\epsilon_{p}(k_{n,p})-\epsilon_F-n\Omega=E$,
 with $\epsilon_p(k)=\frac{k^2}{2 m}-\Delta_p$ being the dispersion of channel $p$ in the wire. 
Here $E\in[-\Omega/2,\Omega/2]$ is the Floquet energy of the scattering states and $\epsilon_F$ is the Fermi-energy of 
the incoming electrons in the lead in the absence of a bias voltage. 
Note that the absence of a time-dependent potential in the leads is crucial in allowing us to consider states with a definite 
Floquet index $n$. Thus in the presence of a periodically time-dependent scattering, in addition to the channel index $p$,
one must label incoming states and outgoing states in the lead 
with a combination $(n,p)$ of the  Floquet index $n$ and the usual channel index $p$.
The sign of the wave-vectors $k_{n,p}$ in each of the leads $L,R$  
determines whether the mode is moving into or out of the junction region $C$.
Since $\epsilon_F$ is the Fermi-energy of the incoming electrons in the leads, the Floquet index $n$ for 
the incoming modes must satisfy $n=0$. We will denote the Floquet index of the out-going 
state by $n'$.

Thus, the solutions to Eq.~\ref{eq:floquet} can be represented conveniently in terms of a scattering matrix (S-matrix) $S_{n' ,p';n=0, p}$ 
of states from the incoming modes with Floquet index $n=0$ in channel $p$ to out-going modes with $n'$ in channel $p'$ 
 at a fixed floquet energy $E$. The S-matrix is defined by the equation 
\begin{equation}
v_{n',p'}^{1/2}\phi_{n' p'}^{(out)}= \sum_{p}\mathcal{S}_{n' p',n=0,p}(\Omega,E)v_{n=0, p}^{1/2}\phi_{n=0, p}^{(in)}
\end{equation}
where $v_{n=0,p}$ are the group velocities in the respective channels. The scattering matrix $S$ is 
unitary as a consequence of particle-number conservation.

\section{Unitarity of the $\mathcal{S}$-matrix and the Landauer formula} 
The current flowing $I$ from the left-lead $L$ to the right lead $R$ in the geometry in Fig.~\ref{fig1} is a function of the Fermi-energy 
$\epsilon_F$ of the incoming modes of the leads and the bias voltage $V_b$.
 In the presence of a bias voltage, $V_b$, the Fermi-energy of the 
incoming electrons in the left lead, $\epsilon_{F,L}$, differs from the Fermi-energy of the right lead $\epsilon_{F,R}=\epsilon_{F,L}-V_b$.
The conductance $\sigma(\epsilon_F)$, at an equilibrium Fermi-energy $\epsilon_F$ is defined as the response of the current $I(\epsilon_F,V_b)$
 to an infinitesimal bias voltage $V_b$ i.e. $\sigma(\epsilon_F)=dI(\epsilon_F,V_b)/dV_b|_{V_b=0}$.
 Therefore the current $I(\epsilon_F,V_b)$ is a 
function of the mean Fermi-level $\epsilon_F$ and the bias voltage $V_b$.

However, to derive the Landauer formula it is convenient to consider the current $I$ as being a function of the
 Fermi-level of the incoming electrons in the left and right lead which are written as 
\begin{align}
&\epsilon_{F,L}=\epsilon_F+V_b/2\\
&\epsilon_{F,R}=\epsilon_F-V_b/2.
\end{align}
Non-interacting electrons coming from the left lead $L$ remain decoupled from electrons coming in from the right lead $R$. Therefore,
 the current  $I(\epsilon_{F,L},\epsilon_{F,R})$ can be separated as $I(\epsilon_{F,L},\epsilon_{F,R})=I_L(\epsilon_{F,L})+I_R(\epsilon_{F,R})$ 
where $I_L(\epsilon_{F,L})$ ($I_R(\epsilon_{F,R})$) is the current carried by states from the left (right) lead L (R) with
 Fermi-level $\epsilon_{F,L}$( $\epsilon_{F,R}$). 
 
In the model for non-interacting electrons considered explicitly here, changing the Fermi-level in either lead is equivalent to 
applying a voltage to that lead. In this case, the mean Fermi-level $\epsilon_F$ in the wire can be easily changed
 by applying a voltage symmetrically (i.e. $\delta\epsilon_{F,L}=\delta\epsilon_{F,R}$) 
to the two leads. For interacting electrons, however, it is difficult to change the net charge density in the wire,
 so a change in the 
average voltage of the leads produces primarily a change in the electrostatic potential of the wire and only a slight change 
in the mean Fermi-energy $\epsilon_F$.
However, if the sample has a side gate in the vicinity of the wire, which will be needed in our proposal to produce the confinement needed 
to create the wire in any case, it may be possible to produce a 
significant change in the electron density in the wire, and thus vary $\epsilon_F$, by applying a large voltage difference 
between the gate and the leads to the wire.
 Although we consider 
explicitly only the non-interacting case, we shall write our final results in terms of $\epsilon_F$ and $V_b$, rather than 
$\epsilon_{F,R}$ and $\epsilon_{F,L}$ in the expectation that they will also be applicable at least qualitatively, for an interacting 
system.

While the S-matrix representation is convenient because it leads to a  simple condition for probability conservation, for transport 
between the left and right leads, it is more convenient to think in terms of the transmission or $\mathcal{T}$-matrix, which is the 
part of the S-matrix defined as 
\begin{equation}
S_{n'p';np}=\left(\begin{array}{cc}\mathcal{R}_{n'p';np}^{(L\rightarrow L)}&\mathcal{T}_{n'p';np}^{(L\rightarrow R)}\\
\mathcal{T}_{n'p';np}^{(R\rightarrow L)}&\mathcal{R}_{n'p';np}^{(R\rightarrow R)}\end{array}\right),
\end{equation}
where $\mathcal{R}$ is the reflection-matrix. The superscripts $L$ and $R$ denote whether we have modes going from left to right 
or vice-versa.
Applying the standard argument used in the derivation of the Landauer formula to the incoming states in the 
left-lead with Floquet energies in the range $E_L=\epsilon_{F,L}-\epsilon_F\in [0,V_b/2]$,
 the part of the transmission matrix $\mathcal{T}_{L\rightarrow R}$ can be used to calculate the current $I_L(\epsilon_{F,L})$ and 
therefore also the derivative  of the dc current $I$  with respect to the left lead chemical potential $\epsilon_{F,L}$, as 
\begin{align}
&\sigma_L(\epsilon_F)=\frac{\partial I}{\partial\epsilon_{F,L}}=\frac{d I_L}{d\epsilon_{F,L}}|_{V_b=0}\nonumber\\
&=G_0\sum_{ p',n',p}|\mathcal{T}_{n' p';n=0 p}^{(L\rightarrow R)}(\Omega,E=0)|^2,\label{floquetlandauer}
\end{align}  
where the channels $p$ summed over for the occupied incoming states in $L$ and $n'$ and $p'$ are, respectively,
 the  Floquet and channel indices of the outgoing modes in $R$.
 The response of the current $I$ with respect to the right chemical potential $\epsilon_{F,R}$, which is defined as 
 $\sigma_R(E)=-\frac{d I}{d\epsilon_{F,R}}$, can be calculated using an expression similar to Eq.~\ref{floquetlandauer}. 
Note the negative sign in the definition of 
$\sigma_R(E)$ accounts for the direction of the current $I$ being from $L$ to $R$.  
Since the current $I$ in static systems responds only to the difference in the chemical 
potential $V_b=\epsilon_{F,L}-\epsilon_{F,R}$, it is convenient to define the conductance 
 across the leads as the response 
of the current to the bias voltage as  
\begin{equation}
\sigma(\epsilon_F)=\frac{d I}{d V_{b}}=\frac{\sigma_R(\epsilon_F)+\sigma_L(\epsilon_F)}{2},\label{sigma}
\end{equation}
where $\epsilon_F$ is held fixed when taking the derivative.

Static systems, in the absence of a bias voltage $V_b=0$, are in equilibrium so that  the total charge in the
 right lead $R$ must be time-independent.
Therefore the current  $I(\epsilon_F,V_b=0)$ vanishes in the absence of a bias voltage. 
However the driven point contact, in addition to carrying a current in response to a bias voltage $V_b$, 
may also carries a pumping current.
The derivative of the pumping current is calculated as  
\begin{equation}
\sigma_P(\epsilon_F)=\frac{d I_P}{d \epsilon_F}=\frac{d I_L}{d \epsilon_{F,L}}+\frac{d I_R}{d \epsilon_{F,R}}=\frac{\sigma_L(\epsilon_F)-\sigma_R(\epsilon_F)}{2},\label{sigmap}
\end{equation}
where $V_b$ is held at $V_b=0$ when taking the derivative.
The absence of a pumping current $I_P=0$ in the static case leads to the constraint 
\begin{equation}
\sigma_L(\epsilon_F)=\sigma_R(\epsilon_F)\textrm{ for }\Omega=0.\label{constraint}
\end{equation} 
The pumping current $I_P(\epsilon_F)=I(\epsilon_F,V_b=0)$ is calculated using the relation 
\begin{equation}
I_P(\epsilon_F)=\int_{0}^{\epsilon_F}d\epsilon \sigma_P(\epsilon),\label{eq:Ip}
\end{equation}
where $\epsilon_F$ is the Fermi-energy at which the current is measured. 

The above discussion has been restricted to zero temperature.
 Finite temperature modifications of the above results are obtained by averaging the above conductances with 
the function $\frac{d f(E+\epsilon_F;\epsilon_F,T)}{dE}$ where $f(E;\epsilon_F,T)$ is the Fermi function with temperature $T$
 and Fermi-energy
 $\epsilon_F$. For the rest of this paper we will assume that the temperature $T$ is significantly smaller than the driving 
frequency so that it does not qualitatively affect our results.

The expression Eq.~\ref{floquetlandauer} is a generalization of the usual Landauer formula, which is obtained
 from Eq.~\ref{floquetlandauer} 
by considering the static limit where the time-dependent potential $V_{p,q}(x)=0$ vanishes, so that the Floquet-index $n$ 
is a conserved quantity and the Floquet index $n'=0$ for the out-going states. In this limit, all
 incoming modes from the left lead $L$ with energy in the range $\epsilon_{F,R}< (E+\epsilon_F)<\epsilon_{F,L}$
 are occupied while all the outgoing modes (in both $L$ and $R$) are occupied according to
 the scattering matrix $\mathcal{T}$.
Since the matrix $\mathcal{T}^{(L\rightarrow R)}$ is only part of the unitary S-matrix, the sum over all outgoing modes $(p')$ of 
\begin{equation}
\sum_{p'\in R,p\in L}|\mathcal{T}_{p'p}^{(L\rightarrow R)}(\Omega=0,E)|^2<\sum_{p'\in R,p}|\mathcal{S}_{p'p}(\Omega=0,E)|^2 =N_{R}
\end{equation}
where $N_R$ is the number of  channels $p'$ in the right lead $R$.
 It follows that the conductance from the left lead is less than $\sigma_L(\epsilon_F)<N_R$.
In particular, in the case where the right lead $R$ has a single channel i.e. $N_R=1$, the conductance is bounded by $\sigma_L(\epsilon_F)<1$.
Using Eq.~\ref{constraint}, it follows that $\sigma_R(\epsilon_F)<1$ 
and hence $\sigma(\epsilon_F)<1$. 

In the case of a time-dependent perturbation, multiple values of the out-going  Floquet-index $n'$ are allowed
 and at least, in principle, the conductance can exceed 1. Such a conductance exceeding unity provides a direct test of the
 accessibility of
 Floquet modes indexed by the photon number 
$n$. Below we discuss one scenario where the conductance can exceed 1. Furthermore, the geometry we will discuss 
will also have a non-vanishing pumping current, even for single parameter pumping.

\section{Schematic system}
Let us now discuss a specific case of the geometry  shown in Fig. 1 where the conductance can exceed the conductance quantum. 
In particular, we assume that the portions $L$ and $C$ of the wire have two transmitting channels each, while the region $R$ has a 
single transmitting channel, so that the static conductivity is bounded above by unity. For simplicity, the 
scattering potential in the central region $C$ is split into a time-dependent but reflection-less 
scattering part with a transmission matrix $\mathcal{T}_F(\Omega,E)$ and a static, but only partially transmitting,  
part with a transmission matrix $\mathcal{T}_0$.
The reflection-less potential with transmission matrix $\mathcal{T}_F(\Omega,E)$ can be approximately realized by 
 a grating potential, which only has  wave-vector components on the scale of $\pm (k_{F,1}-k_{F,2})$.
  The transmission matrix $\mathcal{T}_0$ is realized in the usual way, 
 constricting the quasi-one-dimensional channel adiabatically from a width that supports two 
channels at the Fermi level to a width that supports only one.
 The time-dependent part of the scattering  $\mathcal{T}_F(\Omega,E)$, being  almost purely transmitting, 
only transfers electrons between channels $p=1,2$ in the lead $L$ and central region $C$.
The static scattering $\mathcal{T}_0$ completely reflects in channel $p=2$ and transmits 
channel $p=1$ from $C$ to $R$, so that the  using the Floquet Landauer formula Eq.~\ref{floquetlandauer},
 the left conductance from $L$ to $R$ is 
\begin{equation}
\sigma_L(\epsilon_F)=\sum_{p=1,2;n'}|\mathcal{T}_{F, n' 1;n=0 p}(\Omega,E=0)|^2.\label{sigmaL0}
\end{equation}    
Using the unitarity of the $\mathcal{T}_F$ matrix in Eq.~\ref{sigmaL0}, the term $\sum_{n'}|\mathcal{T}_{F,n' 1; 0 1 }(\Omega,E=0)|^2$
can be replaced by $\sum_{n'}|\mathcal{T}_{F, n' 1;0 1 }(\Omega,E=0)|^2=1-\sum_{n'}|\mathcal{T}_{F,n' 2; 0 1 }(\Omega,E=0)|^2$,
 so that $\sigma_L$ is re-written as 
\begin{align}
&\sigma_L(\epsilon_F)=1\nonumber\\
&+\sum_{n'}|\mathcal{T}_{F,n' 1;0 2 }(\Omega,E=0)|^2-|\mathcal{T}_{F, n' 2;0 1}(\Omega,E=0)|^2.\label{sigmaL}
\end{align} 

The conductance from the right lead is entirely determined by the fact that $\mathcal{T}_0$ transmits channel $1$
 in the left lead perfectly into the right lead and reflects channel $2$. This corresponds to a $\sigma_L(\epsilon_F)=1$ 
for just the transmission matrix $\mathcal{T}_0$.  Since $\mathcal{T}_0$ is the result of 
a time-independent scattering, using Eq.~\ref{constraint}, one obtains $\sigma_R(\epsilon_F)=\sigma_L(\epsilon_F)=1$,
so that the only incident channel i.e. channel $p=1$ must be completely transmitted from right to left. The 
potential for the time-dependent scattering i.e. $\mathcal{T}_F$ is reflectionless and therefore must also 
completely transmit these electrons in channel $p=1$. The combination of these arguments lead to the result that, 
even in the presence of an AC perturbation,  
\begin{equation}
\sigma_R(\epsilon_F)=1.
\end{equation}
Such configurations are characterized by a derivative of the pumping current $\sigma_P$ satisfying
\begin{equation}
\sigma_P(\epsilon_F)=\sigma(\epsilon_F)-1,\label{sigmaP1}
\end{equation}
so that the existence of the pumping current is related to having a conductance $\sigma$ above the Landauer limit $\sigma=1$.

To maximize the conductance $\sigma(\epsilon_F)$ in Eq.~\ref{sigma}, one must maximize $\sigma_L(\epsilon_F)$.
 Since the sum $\sum_{n'}|\mathcal{T}_{F, n' 1;0 2}(\Omega,E=0)|^2 < 1$, the conductance $\sigma_L(\epsilon_F)$ in
 Eq.~\ref{sigmaL} is bounded by unitarity of $\mathcal{T}_{F}$ by $2$. Therefore, if one can 
construct a system where $\sum_{n'}|\mathcal{T}_{F, n' 1; 0 2}(\Omega,E=0)|^2\approx 1$
 and $|\mathcal{T}_{F, n' 2;0 1}(\Omega,E=0)|\approx 0$ then $\sigma_L(\epsilon_F)$ can reach it's maximum value of $2$.
 If $\sigma_L(\epsilon_F)$ reaches its maximum value of 2,
 the conductance $\sigma(\epsilon_F)$ would reach a value of 1.5, which is in excess of the Landauer limit of 
$\sigma=1$ for a single channel wire.
 In addition, there would be a pumping current from the fact the existence of a non-vanishing differential pumping 
current $\sigma_P(\epsilon_F)\sim 0.5$.  
Below, we discuss a realization of $\mathcal{T}_F$ which would produce such a property.

\section{Calculation of $\mathcal{T}_F(\Omega,E=0)$}
\begin{figure}
\centering
\includegraphics[scale=0.35,angle=0]{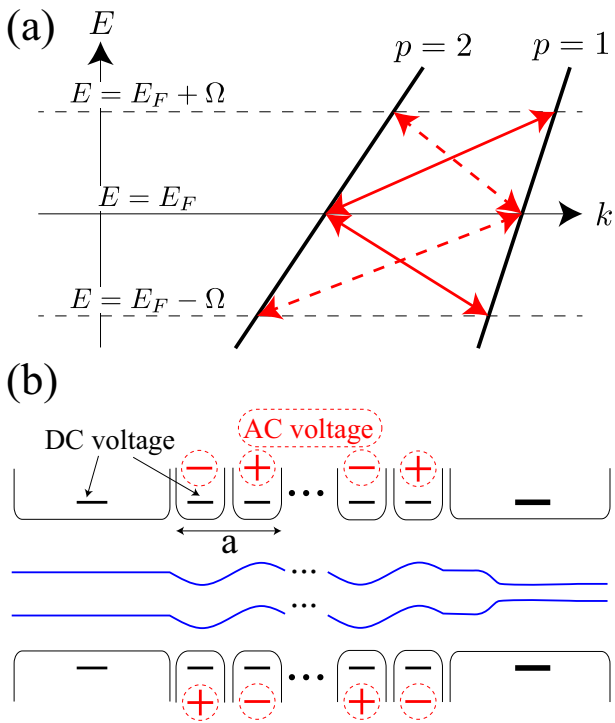}
\caption{ (a)Schematic of the scattering induced by $\mathcal{T}_F(\Omega)$. 
The bandstructure of the two-channel lead $L$ contains two channels $p=1,2$. The time-dependent 
potential producing the scattering $\mathcal{T}_F(\Omega)$ is required to be tuned so that it produces strong transitions 
from channels $p=2$ to $p=1$ along the solid arrows and only weak scattering from $p=1$ to $p=2$ along the dashed arrows.
These transitions are separated by different momenta. 
(b) Nearly periodic grating structure to create the momentum selectivity needed for $\mathcal{T}_F(\Omega)$.
The DC voltages on the left end of the wire are smaller so as to confine a two-channel wire in between the solid lines,
 while at the right end they are larger so as to confine a narrower single-channel wire.
 The grating structure in the middle consists of a set of electrodes 
which are separated from each other so that an alternating AC voltage can be superimposed on them in addition to the DC voltage. 
The alternating AC voltage distorts the potential in the channel in a sinusoidal fashion as shown leading to an effective sinusoidal 
scattering potential between the channels $p=1$ and $p=2$.
}\label{fig2}
\end{figure}

We now discuss a specific model for the time-dependent scattering  $\mathcal{T}_F(\Omega,E)$, which is shown in Fig.~\ref{fig2}, 
 that can lead to the ideal amplitude scattering described in the previous paragraph. The transmission matrix
 $\mathcal{T}_F(\Omega,E)$ can be obtained by solving Eq.~\ref{eq:floquet} written  as 
\begin{align}
&\left(-\frac{\partial_x^2}{2 m}+\Delta_{p'}-\epsilon_F\right) \phi_{n,p'}-(n\Omega+E)\phi_{n, p'}\nonumber\\
&=\sum_p V_{p',p}(x)\phi_{n-1, p}+V^*_{p',p}(x)\phi_{n+1, p},\label{eq:phi}
\end{align}
where we have used the Hamiltonian components in Eq.~\ref{fullH}.
 As mentioned previously, we will choose a potential $V_{p',p}(x)$ for the transmission amplitude 
$\mathcal{T}_F(\Omega,E)$, so that the Fourier components at the back-scattering wave-vectors $2 k_{F,p}$ and $k_{F,1}+k_{F,2}$
 can be neglected.  
Therefore, for frequencies and temperatures much lower than the Fermi energy for each of the channels, we can restrict 
our attention to states near the right moving Fermi points in the two channels with Fermi-wave-vectors $k_{F,1,2}$ in the 
lead $L$. For such states of the Hamiltonian in Eq.~\ref{fullH}, it is convenient to transform the wave-functions as 
 $\phi_{n, p}(x)=e^{i k_{n, p} x}\tilde{\phi}_{n, p}(x)$, and ignore second derivatives of the slowly varying 
(compared to the absolute Fermi-wave-vector $k_{F,1,2}$) wave-function $\tilde{\phi}_{n, p}(x)$ 
 so that Eq.~\ref{eq:phi} simplifies to  
\begin{align}
&\partial_x \tilde{\phi}_{n, p'}=-i v_{F,p'}^{-1}e^{-i k_{n,p'} x}\nonumber\\
&\sum_p [ V_{p',p}(x) e^{i k_{n-1,p} x}\tilde{\phi}_{n-1, p}+V^*_{p',p}(x)e^{i k_{n+1,p} x}\tilde{\phi}_{n+1, p}]\label{f:scatter}
\end{align}
where $k_{n,p}=(k_{F,p}+n\frac{\Omega}{v_{F,p}})$.  
Propagating these equations with initial values at $x=0$ so that $\tilde{\phi}_{n,p}(x=0)$
 is non-zero only for the incoming modes with $n=0$, one can calculate the 
coefficients $\mathcal{T}_{F,n' p';0 p}(\Omega)$ through the equation 
\begin{equation}
\sum_{p}\mathcal{T}_{F,n' p';0 p}(\Omega)\sqrt{v_{F,p}}\phi_{0,p}(x=0)=\sqrt{v_{F,p'}}\phi_{n',p'}(x=L).
\end{equation}

Calculating the scattering matrix $\mathcal{T}_F$ for a weak scattering potential $V$ shown
 in Fig.~\ref{fig2}  using the first order Born approximation yields insight into the mechanism for obtaining a large 
Floquet conductance.
Given that the incident states are at photon number $n=0$, the final states in region $C$, following the time-dependent scattering,
 are written as 
\begin{align}
&\tilde{\phi}_{1,p'}(L)=-i v_{F,p'}^{-1}\int_0^L dx e^{-i k_{1,p'} x}\sum_p V_{p',p}(x)e^{i k_{0,p} x}\tilde{\phi}_{0,p}(0)\nonumber\\
&\tilde{\phi}_{-1,p'}(L)=-i v_{F,p'}^{-1}\int_0^L dx e^{-i k_{-1,p'} x} \sum_p V_{p',p}^*(x) e^{i k_{0,p} x}\tilde{\phi}_{0,p}(0),\label{eq:psi1}
\end{align}
for $p',p=1,2$. 
Within the first order Born approximation, only the scattering processes shown by arrows in Fig.~\ref{fig2}(a), 
which absorb or emit energy, contribute to the conductance in Eq.~\ref{sigmaL}.
 In order to achieve the required $\mathcal{T}_F(\Omega)$, which produces 
a  conductance $\sigma(\epsilon_F)>1$  and a finite pumping current, we will need to ensure that the scattering amplitude for 
the solid arrows in Fig.~\ref{fig2}(a) are larger than the one shown in the dashed arrows.

One way to realize the required constraints on the scattering amplitude is to use the combination of confinement 
and grating potentials shown in Fig.~\ref{fig2}(b). The electrodes confine the 2DEG in the middle into a quantum wire, 
while the grating structure allows one to apply a time-dependent potential, which is periodic in space over a certain 
length-scale. One can expect such a grating structure to allow us to select specific momentum transfers so that 
the transitions along the solid arrows are preferred over the dashed ones. Since we will choose the selected momentum 
transfer of the grating to be of order $\delta k_F=k_{F,1}-k_{F,2}$, which is different from the 
back-scattering wave-vectors, the back-scattering rate can be shown to be suppressed. For simplicity, we will assume that 
 the grating potential is applied 
with opposite polarities on the two sides of the wire, while the wire is reflection symmetric so that the mode-wavefunctions 
for the channels $p=1$ and $p=2$ have opposite parities. Under these conditions, the applied time-dependent potential $V(x,y)$
 leading to the matrix-elements $V_{p',p}(x)$ leads to off-diagonal matrix element between the channels $p=1$ and $p=2$ so that
 $V_{p',p}(x)=(1-\delta_{p',p})f(x)$. Here $ f(x)=\int dq dk dy \phi_{1,k+q/2}^*(y)V(x,y)e^{i q x}\phi_{2,k-q/2}(y)$ 
and $\phi_{p=1,2;k}(y)$ are the possibly momentum dependent transverse wave-functions of the modes in the $y$ direction.

The periodic nature of the grating structure in Fig.~\ref{fig2}(b) that generates the potential $f(x)$,
 allows us to write $f(x)=\sum_{j=0}^{N-1} g(x-j a)$ where $g(x)$ is the potential at the wire generated by one neighboring 
pair of electrodes in the grating in Fig.~\ref{fig2}(b) 
with lattice constant $a$.
The relevant integral over $x$ of $f(x)$ is written as 
\begin{equation}
f(q)=\int_{0}^L dx f(x) e^{i q x}\approx g(q)e^{i q a (N-1)/2}\frac{\sin{(N q a/2)}}{\sin{(q a/2)}}\label{fq}
\end{equation}
 where $g(q)$ is the fourier transform of $g(x)$ and we have assumed that $g(x)$ is such that $f(x)\approx 0$ outside the interval $[0,L]$.
Because the potential $g(x)$ is constructed using electrodes with opposite signs, $g(q=0)$ vanishes at $q=0$ and the 
largest resonance (of order $N g(q=\frac{2\pi}{a})$) in $f(q)$ occurs at $q a\sim 2\pi$. The other resonances occuring 
at higher harmonics of this fundamental wave-length are expected to be suppressed for electrodes placed at a distance 
further than $a$ from the wire. The width of the peak in $f(q)$ is expected to be of order $\delta q\sim \frac{4\pi }{N a}$.

Performing the integrals in Eq.~\ref{eq:psi1}, we obtain
the Floquet scattering matrices relevant to the calculation of $\sigma_L(\epsilon_F)$
 in Eq.~\ref{sigmaL}, which are then written as 
\begin{align}
&\mathcal{T}_{F,1,1;0,2}=-i v_{F,1}^{-1/2} v_{F,2}^{-1/2} f(k_{0,2}-k_{1,1})\nonumber\\
&\mathcal{T}_{F,-1,1;0,2}=-i v_{F,1}^{-1/2}v_{F,2}^{-1/2} f^*(-(k_{0,2}-k_{-1,1}))\nonumber\\
&\mathcal{T}_{F,1,2;0,1}=-i v_{F,1}^{-1/2} v_{F,2}^{-1/2} f(k_{0,1}-k_{1,2})\nonumber\\
&\mathcal{T}_{F,-1,2;0,1}=-i v_{F,1}^{-1/2}v_{F,2}^{-1/2}  f^*(-(k_{0,1}-k_{-1,2})),\label{scattering_amp}
\end{align}
where intra-band scattering processes have been assumed to be suppressed because of symmetry.
The upper pair of scattering elements correspond to the solid arrows in Fig.~\ref{fig2}(a) and the lower pair
of scattering elements corresponds to the dashed arrows. It follows from Eq.~\ref{sigmaL} that the conductance 
$\sigma_L(\epsilon_F)>1$, whenever $\sum_{n'}|\mathcal{T}_{F,n' 1;0 2 }(\Omega,E=0)|^2-|\mathcal{T}_{F, n' 2;0 1}(\Omega,E=0)|^2>0$ i.e.
 the prcesses associated with the solid arrows in Fig.~\ref{fig2}(a) dominate over the dashed arrows.

The requirement that the scattering amplitudes shown by the solid arrows in Fig.~\ref{fig2} are stronger in magnitude 
than the dashed arrows can be satisfied using the momentum dependence of the scattering amplitude
 from the potentials $V_{p',p}(x)$.
This translates into the fourier transform $f(q)$ of the $x$-dependence of the potential having a resonance of width $\delta q$ 
at a momentum corresponding to one of the solid arrows, say the upper solid arrow.
 To obtain large conductance $\sigma(\epsilon_F)>1$, the width of the 
resonance in $f(q)$ has to ensure that the scattering associated with the dashed arrows in Fig.~\ref{fig2} are suppressed.  
The process associated with the other solid arrow need not be suppressed.
The Fourier transform $f(q)$ for the potential in Eq.~\ref{fq}
 has a peak of height order $\sim N$ and width of order $\delta q\sim \frac{4\pi }{N a}$ near $q a=2 \pi$.
 By tuning a gate voltage, and correspondingly the mean-Fermi energy $\epsilon_F$ to $\epsilon_F=\epsilon_{F,0}$,
the Fermi wave-vectors $k_{F,1}$ and $k_{F,2}$ of the incident right moving electrons 
in $L$ can be tuned so that the condition 
\begin{equation}
k_{F,1}-k_{F,2}=\frac{2\pi}{a}\label{epsF0}
\end{equation}
is satisfied and  all the scattering processes 
represented by the arrows in Fig.~\ref{fig2}, in the limit of small frequency $\Omega\rightarrow 0$, 
can be driven by the scattering potential $V_{p',p}(x)$, whose Fourier transform has a peak at the wave-vector $q=\frac{2\pi}{a}$.
Changing the Fermi-energy $\epsilon_F$ by $\delta\epsilon_F$ changes $k_{F,p}$ by different 
rates $\delta k_{F,p}=\delta\epsilon_F/v_{F,p}$ for the channels $p=1,2$. Therefore one expects to be able to tune the Fermi-energy 
to be able to satisfy the condition in Eq.~\ref{epsF0}.
At finite frequency $\Omega$, which is still much smaller than $\epsilon_F$, the outgoing wave-vectors $k_{n,p}$ associated with the 
different Floquet indices $n$ are split from $k_p=k_{F,p}$ by amounts proportional to the frequency and 
are given by the relations $k_{F,p}=k_{0,p}=k_{-1,p}+\frac{\Omega}{v_{F,p}}=k_{1,p}-\frac{\Omega}{v_{F,p}}$ 
for the channels $p=1,2$.
The frequency splitting of the wave-vector is crucial to be able achieve the regime where the process associate with the 
upper solid arrow in  Fig.~\ref{fig2}(a) can be made large, while keeping the dashed arrows small. 
The largest possible conductance, $\sigma$, should be obtained when the Fermi-energy $\epsilon_F=\epsilon_{F,0}+\delta\epsilon_F$ 
is tuned to satisfy 
the condition 
\begin{equation}
(k_{1,1}-k_{0,2})a= \left(k_{F,1}-k_{F,2}+\frac{\Omega}{v_{F,1}}\right)a=2\pi,
\end{equation}
which ensures that the upper solid arrow in Fig.~\ref{fig2}(a) dominate the conductance in the small $\delta q$ limit.
In fact, we expect the transfer associated with the  upper solid arrow in Fig.~\ref{fig2}(a) to be large 
as long as 
\begin{equation}
\left|\left(k_{F,1}-k_{F,2}+\frac{\Omega}{v_{F,1}}\right)-\frac{2\pi}{a}\right|=\left|\left(\frac{\delta\epsilon_F+\Omega}{v_{F,1}}-\frac{\delta\epsilon_F}{v_{F,2}}\right)\right|<\frac{2\pi }{N a}.\label{resonance}
\end{equation}
In addition, we require the amplitude for the processes associated with the  dashed arrows in Fig.~\ref{fig2}(a) to be 
small so that 
\begin{align}
&(k_{0,1}-k_{1,2})= \left(k_{F,1}-k_{F,2}+\frac{\Omega}{v_{F,2}}\right)>\frac{2\pi}{a}+\frac{2\pi}{N a}\\
&(k_{0,1}-k_{-1,2})= \left(k_{F,1}-k_{F,2}-\frac{\Omega}{v_{F,2}}\right)<\frac{2\pi}{a}-\frac{2\pi}{N a}.\label{constraints}
\end{align}
These conditions together with the resonance condition Eq.~\ref{resonance} are satisfied if the frequency $\Omega$ satisfies the 
condition  
\begin{equation}
\Omega\left(\frac{1}{v_{F,2}}-\frac{1}{v_{F,1}}\right)>\frac{4\pi}{Na}.
\end{equation} 
This inequality can also be viewed as a constraint on the total length of the grating structure 
\begin{equation}
L=N a\gtrsim \frac{4\pi v_{F,1}v_{F,2}}{\Omega (v_{F,1}-v_{F,2})}.\label{Na}
\end{equation} 
When the parameters of the system satisfy Eq.~\ref{Na} and the Fermi-energy is tuned to satisfy Eq.~\ref{resonance}, we expect the upper 
solid arrow in Fig.~\ref{fig2} to be dominant and using Eq.~\ref{sigmaL} it follows that $\sigma_L(\epsilon_F)>1$.
Using Eq.~\ref{sigma} and Eq.~\ref{sigmap} leads to a conductivity $\sigma(\epsilon_F)>1$, which is  above the
 Landauer limit and a finite pumping 
current $\sigma_P(\epsilon_F)>0$. 

The pumping current $I_P(\epsilon_F)$ at a Fermi-energy $\epsilon_F$ is obtained from $\sigma_P(\epsilon)$ using Eq.~\ref{eq:Ip}.
In principle, evaluating the pumping current $I_P(\epsilon_F)$ requires calculation of $\sigma_P(\epsilon_F)$ for the entire range 
of Fermi-energies.
However, from physical considerations, one expects a significant effect of the time-dependent scattering process, which is what leads 
to the pumping current to occur only when the Fermi-energy is tuned so that the time-dependent scattering is off-resonant.
Therefore $I_P(\epsilon_F)$ should vanish when the Fermi-energy is below the range implied by Eq.~\ref{resonance}. 
Thus, combinging the resonance conditions and constraints in Eqs.~\ref{resonance},\ref{constraints}, $I_P(\epsilon_F)$
 may be estimated by integrating $\sigma_P(\epsilon)$ 
over a range of Fermi-energies $\epsilon_F=\epsilon_{F,0}+\delta\epsilon_F$ around $\epsilon_{F,0}$ satisfying 
\begin{equation}
\frac{\Omega}{v_{F,1}}-\frac{2\pi}{N a}<\delta\epsilon_F\frac{v_{F,1}-v_{F,2}}{v_{F,1}v_{F,2}}<\frac{\Omega}{v_{F,1}}+\frac{2\pi}{N a}.
\end{equation}
The corresponding range of $\delta\epsilon_F$ satisfies 
\begin{equation}
|\delta\epsilon_{F,max}-\delta\epsilon_{F,min}|<\frac{4\pi}{N a}\frac{v_{F,1}v_{F,2}}{v_{F,1}-v_{F,2}}<\Omega.\label{epsrange}
\end{equation}
Given that $\sigma_L(\epsilon_F)<2$ in our geometry, we expect the pumping current $I_P(\epsilon)$ to be bounded by and of order 
$I_P\sim\frac{\Omega}{2}G_0$ for an optimally chosen set of parameters,
 where $G_0=2 e^2/h$ is the quantum of conductance. Since the range of Fermi-energies we have been discussing 
are relevant only for the Fermi-energy of the left-lead $\epsilon_{F,L}$, the bound Eq.~\ref{epsrange} also applies 
to the maximum bias voltage over which the linear response conductance $\sigma(\epsilon_F)$ is reasonably accurate. Thus, 
the bound on the pumping current also serves as an estimate of the excess current signal that one expects to see over the Landauer 
limit. Of course, the discussion of the magnitudes of the currents in this paragraph are rather qualitative and provide only 
order of magnitude estimates. The conductance $\sigma(\epsilon_F)$ and the pumping current in a bias window $I_P$
 can be calculated with quantitative accuracy by numerically integrating the equations Eq.~\ref{eq:phi}.
 However, the magnitudes of the currents discussed in this paragraph will be qualitatively verified by the numerical results presented 
later in the paper.

Let us now discuss a choice for the various parameters in our geometry that may be reasonable to realize in GaAS 2DEG experiments. 
We choose a lattice parameter for the grating to be $a\sim 50$nm. Using Eq.~\ref{resonance},
 the choice for the lattice parameter $a$ sets $k_{F,1}-k_{F,2}\sim \frac{2\pi}{a}= 12.56\times 10^{-2}$ nm$^{-1}$.
Taking a nanowire of width $\sim 35$ nm,~\cite{amir} leads to 
to a linear electron-density of $n_{1D}\sim \frac{6.3}{a}=0.1/$ nm, which corresponds to a two-dimensional 
 electron density of $3.6\times 10^{11}/$ cm$^2$.~\cite{amir} Taking the effective mass of GaAs to 
be $m=0.067 m_e$, we obtain the Fermi-velocities to be $v_{F,1}=4.4 v_{F,2}=2.8\times 10^5$ m/s. Choosing a frequency of
 $\frac{1}{2\pi}\Omega=40$ GHz,
the required length of the grating would be $L\sim 4\,\mu$m. To achieve this length of grating requires $N=L/a=80$ repetitions 
of the unit potential.

\begin{figure}
\centering
\includegraphics[scale=0.35,angle=0]{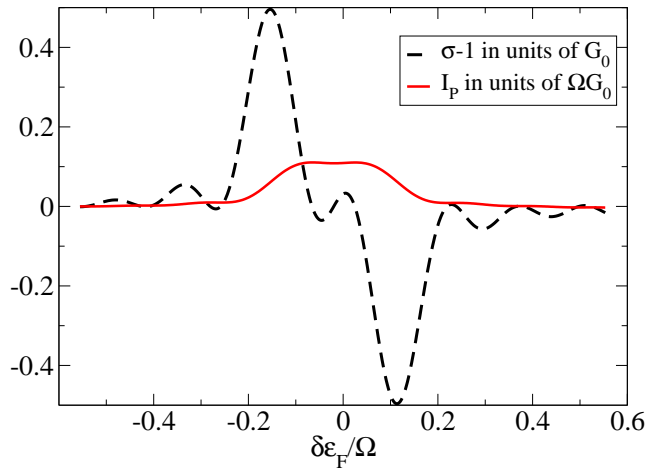}
\caption{
The  transport conductance $\sigma(\delta\epsilon_F+\epsilon_{F,0})-1$ in excess of the Landauer value of unity and
 also the pumping current $I_P(\delta\epsilon_F+\epsilon_{F,0})$ as a function 
of the deviation of the Fermi-energy, $\delta\epsilon_F=\epsilon_F-\epsilon_{F,0}$, 
from a reference Fermi-energy $\epsilon_{F,0}$ defined in Eq.~\ref{epsF0} in the text.
 The deviation of the Fermi-energy $\delta\epsilon_F$ is in units of the driving frequency $\Omega$, while 
the conductance $\sigma(\epsilon_F)-1$ is in units of the conductance quantum $G_0=2 e^2/h$.
 The pumping current is in units of $\Omega G_0$.
 The grating structure is taken to contain $N=80$ cycles of a sinusoidal potential and the total
 length is taken to be $L=N a\sim 4\,\mu$m for a frequency  $(\Omega/2\pi)\sim 40$ GHz
 corresponding to a voltage of $150\,\mu$eV. The applied electric field,
 which is chosen to be $1.5\times 10^2\,$ V/cm (for a channel 
width of 35 $nm$) so as to obtain the maximum conductivity.
}\label{fig3}
\end{figure}

The physics discussed in the preceeding paragraphs, which was based on the Born approximation, can be verified by
numerically solving 
the transmission problem corresponding to the Hamiltonian in Eq.~\ref{eq:phi}  using the transmission matrix 
technique.  
 The results for $\sigma(\epsilon_F)-1$ and $\sigma_P(\epsilon_F)$ are plotted in Fig.~\ref{fig3} and show peaks 
as a function of energy $E$. These peaks correspond to the resonance in Eq.~\ref{resonance}.
Similar peaks occur for the other resonance condition corresponding to the lower solid arrow in Fig.~\ref{fig2}(a).
 One finds that for a time-dependent electric field of $\mathcal{E}\sim 1.6\times 10^2$ V/cm (corresponding to an amplitude 
$\mathcal{A}=\textrm{max}_x |f(x)|\sim 100\,\mu$eV),   a two-terminal DC conductance beyond the Landauer bound of 
$\sigma(\epsilon_F)\sim 1.5\, G_0$  and a differential pumping current of $\sigma_P(\epsilon_F)\sim 0.5\, G_0$ 
 over a range of voltage $\Delta E\sim \Omega$. Therefore the 
total pumping current are of order $I\sim 0.5\Omega\,G_0$ 
where $G_0=2 e^2/h$ is the conductance quantum. The pumping current and transport current are found to be of order $I\sim 1.2$ nA 
for a pumping frequency $(\Omega/2\pi)\sim 40$ GHz corresponding to a voltage of $150\,\mu$eV.

\section{Adiabatic pumping limit}
In the previous sections we have shown that in specific regimes of frequency $\Omega$, length of grating structure $L=N a$ 
and electric-field amplitude $\mathcal{E}$, one can obtain a finite pumping current. 
However, we also argued that, the pumping current must vanish in the absence of a time-dependent perturbation.
The pumping current $I_P(\epsilon_F)$ can be shown to vanish within the Born approximation, in 
the limit of vanishing time-dependent drive-amplitude $\mathcal{E}$, by combining  Eqns.~\ref{sigmap},~\ref{sigmaL} and ~\ref{scattering_amp}.
An alternative approach to obtaining a static system is to reduce the driving frequency $\Omega$ to zero, while keeping   
 the geometric parameters of the system such as the length of grating $L$ and also the electric-field amplitude $\mathcal{E}$ fixed.
Strictly speaking, this limit, which is referred to as the adiabatic limit, may be different from the static limit. 
However, formally the result from the previous section shows that $I_P\propto \Omega$ so that 
 the pumping current, $I_P$, vanishes in the adibatic limit (i.e. $\Omega\rightarrow 0$) consistent with our 
expectations. This result must be considered a formal coincidence, since for fixed grating length $L$, the
 limit $\Omega\rightarrow 0$ would violate the constraint in Eq.~\ref{Na}, so that one cannot simply take the limit $\Omega\rightarrow 0$ 
in the final result from the previous section.

To calculate the adiabatic limit for the current in the Born approximation, we expand the pumping current $I_P$ in powers of $\Omega$, 
for small electric fields $\mathcal{E}$, while 
the geometric parameters are held fixed. The Floquet scattering equations Eq.~\ref{f:scatter} are analytic in both $\Omega$ and $\mathcal{E}$ 
for fixed $L$ so that the limit $\Omega\rightarrow 0$ commutes with the limit $\mathcal{E}\rightarrow 0$.
 Therefore the leading order in $\mathcal{E}$ contribution to the adiabatic 
pumping current is calculated by  substituting the Born scattering amplitudes from Eq.~\ref{scattering_amp}
 into the conductance equations Eq.~\ref{sigmaL} 
and expanding to lowest order in the frequency $\Omega$. Using this procedure, we find that 
 the derivative of the pumping current is given by  
\begin{align}
&\sigma_P(\epsilon_F)\approx \Omega  v_{F,1}^{-2}v_{F,2}^{-2}(v_{F,1}-v_{F,2})\frac{d}{d k}[|f(k)|^2-|f(-k)|^2]|_{k=\delta k_F}\nonumber\\
&-\frac{\Omega^2}{2}  v_{F,1}^{-3}v_{F,2}^{-3}(v_{F,1}^2-v_{F,2}^2)\frac{d^2}{d k^2}[|f(k)|^2+|f(-k)|^2]|_{k=\delta k_F},\label{sigmaPad}
\end{align} 
which vanishes as  $\Omega\rightarrow 0$. Here $\delta k_F=k_{F,1}-k_{F,2}$ and the fourier transform of the 
drive potential $f(k)\propto \mathcal{E}$. The frequency-dependence for the pumping current, $I_P(\epsilon_F)$, 
which can be calculated using Eq.~\ref{eq:Ip}, is found to be similar to  Eq.~\ref{sigmaPad}.  

The time-dependent driving potential in Eq.~\ref{hpq} can be written as 
\begin{equation}
\{V_{p,q}(x)+V_{q,p}^*(x)\}\cos{(\Omega t)}+i\{V_{p,q}(x)-V_{q,p}^*(x)\}\sin{(\Omega t)}.
\end{equation}
The above driving potential can be re-written as 
\begin{equation}
2 [\{f_R(x)\}\cos{(\Omega t)}+f_I(x)\sin{(\Omega t)}]
\end{equation}
where $f_R(x)=\frac{f(x)+f^*(x)}{2}$ and  $f_I(x)=\frac{f(x)-f^*(x)}{2 i}$ are the real and 
imaginary parts of the function $f(x)$. If a  time-dependent 
potential with the same phase is applied to all the electrodes in Fig. ~\ref{fig2},  $f_R(x)$ and $f_I(x)$ 
are proportional to each other so that effectively the spatially varying part of the potential $f(x)$ is real and $f(k)=f^*(-k)$.
This case is referred to as single-parameter pumping and is characterized by the constraint 
$\frac{d}{d k}[|f(k)|^2-|f(-k)|^2]=0$. Using Eq.~\ref{sigmaPad}, it follows that the pumping current 
in the adiabatic limit for our set-up scales as $I_P\propto \Omega^2$.
This scaling is qualitatively distinct from the numerical results presented
 in Fig.~\ref{fig3}, where we found that $I_P \propto \Omega$. 
Our adiabatic analysis suggests that the linear scaling of the pumping current $I_P\propto\Omega$ must cross over to an $I_P\propto\Omega^2$ 
 for sufficiently small $\Omega$, i.e. close to the adiabatic limit. 
Since the result $I_P\propto\Omega$ is expected to hold only for frequencies $\Omega\gtrsim v_F/L$ so that the
 constraint on the grating length $L$ given in Eq.~\ref{Na} is satisfied, one 
expects this cross-over to occur at $\Omega\sim v_F/L$.

The vanishing of the linear-order in $\Omega$ pumping current, in the lowest-order Born approximation, is consistent with the 
low-frequency (i.e. $\Omega\rightarrow 0$) scaling of the pumping current from previous studies.\cite{brouwer}
For general systems, the zero-bias current $I_P$ in the limit of small driving frequency
 $\Omega$ can be proportional to $I_P\propto\Omega$ so that the drive pumps a fixed amount of charge per cycle of the potential.
 This is the reason for referring to the zero-bias current as a pumping current.
 The charge pumped per cycle ~\cite{brouwer}  
\begin{equation}
Q_P=\lim_{\Omega\rightarrow 0}\frac{I_P(\Omega)}{\Omega}
\end{equation}
 is related to a Berry-phase of the ground-state acquired over a cycle of the adiabatic pump.
The existence of a non-zero value for such a Berry phase requires the presence of at least two independent parameters in the driving
 potential so that the adiabatic drive can generate a loop in parameter space.
The driving potential discussed so far in our set-up contains only a single-parameter corresponding to linearly 
polarized radiation. Therefore, based on previous results in the literature, one expects 
 that the pumping charge $Q_P$ 
vanishes in the low-frequency (i.e. small $\Omega$) limit. Given the analytic nature of the scattering equations Eq.~\ref{f:scatter},
 the pumping current in the case of single-parameter pumping is expected to scale as 
$I_P\propto \Omega^2$ as verified to $o(\mathcal{E}^2)$ by our Born approximation result. 

The results obtained on adiabatic pumping in previous work,~\cite{brouwer} together with Eq.~\ref{sigmaPad} suggest that adiabatic 
pumping with a non-vanishing linear order in $\Omega$ term $I_P\sim Q_P\Omega$ may still be 
possible even with significantly shorter structures provided a two-parameter driving potential is 
used. Such a potential may be implemented in the quantum wire set-up shown in Fig.~\ref{fig1} by making the potential 
$f(x)$ complex. Physically, a complex value of $f(x)$ is obtained by applying an $x$-dependent time-delay in the radiation-induced 
potential to the various electrodes used to generate the 
potential $f(x)$. It follows from Eq.~\ref{sigmaPad} that a pumping current proportional to $\Omega$ can be 
obtained by replacing the long-grating structure by a pair of bump potentials $f(x)=e^{-i\phi} g(x-a)+e^{i\phi}g(x+a)$ so that
 $|f(k)|=|g(k)\cos{(k a+\phi)}|\neq |f^*(-k)|$. However, since $\sigma_P(\epsilon_F)\sim \sigma(\epsilon_F)-1\propto \Omega$, 
the resulting deviation of the conductance from the Landauer limit is expected to be small. Therefore, the advantage 
in terms of conductance excess over the Landauer limit is not expected to be significantly enhanced by choosing a two-parameter
driving as opposed to a single-parameter drive.

\section{Conclusion}
In this paper we have shown that the additional degrees of freedom, i.e. the Floquet-index associated with a periodically driven 
system can lead to a conductance, which can exceed the number of channels in a quantum wire. For time-independent systems 
general arguments based on unitarity can be used to bound the conductance by the number of channels in the wire. The 
proposed set-up consists of a two-channel wire connected to a single-channel wire through a time-dependent grating 
potential. For a gating structure of length $L=N a\sim 4\,\mu$m in a GaAs 2DEG (seen in Fig.~\ref{fig3}) a peak in the conductance beyond the 
Landauer limit and an oscillatory pump current as a function of gate voltage $\mu_{av}$ are calculated.  
 Therefore, the Floquet-index manifests itself as an additional channel, which can carry a measurable current of the order of 1.2 nA.
 Furthermore, even at zero-bias the system carries a pumping current proportional to the applied frequency $\Omega$ even for a single-parameter 
drive. However, the lower bound on $\Omega$ set by Eq.~\ref{Na} ensures that this result does not contradict the previous work 
interpreting the pumping current as the Berry-phase in a two-parameter space, which predicts that the pumping current vanishes 
for single-parameter pumping in the adiabatic limit.\cite{brouwer} 

We thank Assa Auerbach, Dimitri Feldman, Yuval Gefen and Michael Moskalets for valuable discussions and comments.
J.S. thanks the Harvard Quantum Optics center for support. This work was supported in part by NSF grant DMR 0906475.
We also acknowledge the support from
Army Research Office with funding from the DARPA
OLE program, Harvard-MIT CUA, NSF Grant No.
DMR-07-05472, AFOSR Quantum Simulation MURI,
the ARO-MURI on Atomtronics.


\begin{thebibliography}{99}
\expandafter\ifx\csname natexlab\endcsname\relax

\fi
\expandafter\ifx\csname bibnamefont\endcsname\relax

\fi
\expandafter\ifx\csname bibfnamefont\endcsname\relax

\fi
\expandafter\ifx\csname citenamefont\endcsname\relax

\fi
\expandafter\ifx\csname url\endcsname\relax

\fi
\expandafter\ifx\csname urlprefix\endcsname\relax

\fi
\providecommand{\bibinfo}[2]{#2} \providecommand{\eprint}[2][]{\url{#2}}

\bibitem{tunneling}
F. Grossmann, T. Dittrich, P. Jung, P. Hanggi,  Phys. Rev. Lett. \textbf{67}  516 (1991);
 M. Holthaus, Phys. Rev. Lett. \textbf{69} 351  (1992).

\bibitem{patunneling}
 T. H. Oosterkamp, T. Fujisawa, W. G. van der Wiel, K. Ishibashi, R. V.
Hijman, S. Tarucha, L. P. Kouwenhoven, , Nature 395,  873 (1998);
 G. Platero, R. Aguado,  Phys. Rep. 395,  1 (2004).

\bibitem{adpumping}
 D. J. Thouless,  Phys. Rev. B \textbf{27}, 6083 (1983)
; B. L. Altshuler, L. I. Glazman, Science, \textbf{283}, 1864 (1999) ;
 M. Switkes, C. M. Marcus, K. Campman, A. C. Gossard, Science \textbf{283}, 1905 (1999) .

\bibitem{torres} L. E. F. Foa Torres, Phys. Rev. B, \textbf{72}, 245339 (2005).

\bibitem{brouwer} P. W. Brouwer, Phys. Rev. B \textbf{58}, R 10135 (1998).

\bibitem{nadpumping} M. Wagner, F. Sols,  Phys. Rev. Lett. \textbf{83},  4377(1999);
Y. Levinson, O. Entin-Wohlman, P. Wolfle, Phys. Rev. Lett \textbf{85}, 634  (2000).

\bibitem{lindner}N. H. Lindner, G. Rafael and V. Galitski, Nat. Phys. 7 490
(2011)

\bibitem{takuya}
    T. Kitagawa, T. Oka, A. Brataas, L. Fu, E. Demler, arXiv:1104.4636 (2011).

\bibitem{takuya1}
T. Kitagawa, E. Berg, M. Rudner, E. Demler, Phys. Rev. B 82, 235114 (2010) 

\bibitem{sambe}
H. Sambe, Phys. Rev. A \textbf{7},  2203 (1973).


\bibitem{zudov}  M. A. Zudov, et. al., Phys. Rev. B \textbf{64}, 201311 (2001); P. D. Ye, et.
al., Appl. Phys. Lett. \textbf{79}, 2193 (2001); R. G. Mani,  et al., Nature \textbf{420},
646 (2002); M. A. Zudov et al., Phys. Rev. Lett. \textbf{90}, 046807 (2003).

\bibitem{mirlin} Several alternate mechanisms have been proposed to explain these
experiments.  See, for example, I.M. Dimitriev, A. D. Mirlin, and D. G. Polyakov,
Phys. Rev. B \textbf{75}, 245320 (2007);   A. Auerbach and V. V. Pai, Phys. Rev. B
\textbf{76}, 205318, (2007);  S. A. Mikhailov, Phys. Rev. B \textbf{83}, 155303 (2011) ; and references
therein.


\bibitem{landauer1}
R. Landauer, Phil. Mag. \textbf{21},  863 (1970).

\bibitem{landauer2} E. N. Economou and C. M. Soukoulis, Phys. Rev. Lett. \textbf{46}, 618 (1981);\textit{ibid}. \textbf{47} 972 (1981).

\bibitem{landauer3} D. S. Fisher, P. A. Lee, Phys. Rev. B, \textbf{23}, 6851 (1981).





\bibitem{floquet_transport}
A. P. Jauho, N. S. Wingreen and Y. Meir, Phys. Rev. B 50, 5528 (1994);
S. Kohler, J. Lehmann, P. Hanggi, Phys. Rep. \textbf{406}, 379 (2005);
 M. Moskalets, M. Buttiker, Phys. Rev. B 66 205320 (2002);
 L. Arrachea, M. Moskalets, Phys. Rev. B 74 245322 (2006);
 Hernan L. Calvo, et al, Appl. Phys. Lett. 98, 232103 (2011);
 D. Martinez, R.Molina, B. Hu, 78 045428 (2008).


\bibitem{feldman}
D. E. Feldman and Y. Gefen, Phys. Rev. B \textbf{67}, 115337 (2003).

\bibitem{kaestner}
B. Kaestner,
V. Kashcheyevs,
S. Amakawa,
M. D. Blumenthal,
L. Li,
T. J. B. M. Janssen,
G. Hein, K. Pierz, T. Weimann, U. Siegner, and H. W. Schumacher, Phys. Rev. B \textbf{77}, 153301 (2008).

\bibitem{amir}
G. Barak, H. Steinberg, L. N. Pheiffer, K. W. West, L. Glazman, F. von Oppen, A. Yacoby, 
 Nature Physics \textbf{6}, 489 (2010).

\end{thebibliography}
\end{document}